\documentclass[12pt]{article}

\renewcommand{\theequation}{\arabic{section}.\arabic{equation}}

\def\hybrid{\topmargin -20pt    \oddsidemargin 0pt
        \headheight 0pt \headsep 0pt
        \textwidth 6.25in       
        \textheight 9.5in       
        \marginparwidth .875in
        \parskip 5pt plus 1pt   \jot = 1.5ex}

\hybrid

\catcode`\@=11

\def\marginnote#1{}

\newcount\hour
\newcount\minute
\newtoks\amorpm
\hour=\time\divide\hour by60
\minute=\time{\multiply\hour by60 \global\advance\minute by-\hour}
\edef\standardtime{{\ifnum\hour<12 \global\amorpm={am}%
        \else\global\amorpm={pm}\advance\hour by-12 \fi
        \ifnum\hour=0 \hour=12 \fi
        \number\hour:\ifnum\minute<10 0\fi\number\minute\the\amorpm}}
\edef\militarytime{\number\hour:\ifnum\minute<10 0\fi\number\minute}

\def\draftlabel#1{{\@bsphack\if@filesw {\let\thepage\relax
   \xdef\@gtempa{\write\@auxout{\string
      \newlabel{#1}{{\@currentlabel}{\thepage}}}}}\@gtempa
   \if@nobreak \ifvmode\nobreak\fi\fi\fi\@esphack}
        \gdef\@eqnlabel{#1}}
\def\@eqnlabel{}
\def\@vacuum{}
\def\draftmarginnote#1{\marginpar{\raggedright\scriptsize\tt#1}}

\def\draft{\oddsidemargin -.5truein
        \def\@oddfoot{\sl preliminary draft \hfil
        \rm\thepage\hfil\sl\today\quad\militarytime}
        \let\@evenfoot\@oddfoot \overfullrule 3pt
        \let\label=\draftlabel
        \let\marginnote=\draftmarginnote
   \def\@eqnnum{(\theequation)\rlap{\kern\marginparsep\tt\@eqnlabel}%
\global\let\@eqnlabel\@vacuum}  }

\def\preprint{\twocolumn\sloppy\flushbottom\parindent 2em
        \leftmargini 2em\leftmarginv .5em\leftmarginvi .5em
        \oddsidemargin -.5in    \evensidemargin -.5in
        \columnsep .4in \footheight 0pt
        \textwidth 10.in        \topmargin  -.4in
        \headheight 12pt \topskip .4in
        \textheight 6.9in \footskip 0pt
        \def\@oddhead{\thepage\hfil\addtocounter{page}{1}\thepage}
        \let\@evenhead\@oddhead \def\@oddfoot{} \def\@evenfoot{} }

\def\numberbysection{\@addtoreset{equation}{section}
        \def\theequation{\thesection.\arabic{equation}}}

\def\underline#1{\relax\ifmmode\@@underline#1\else
        $\@@underline{\hbox{#1}}$\relax\fi}

\def\titlepage{\@restonecolfalse\if@twocolumn\@restonecoltrue\onecolumn
     \else \newpage \fi \thispagestyle{empty}\c@page\z@
        \def\thefootnote{\fnsymbol{footnote}} }

\def\endtitlepage{\if@restonecol\twocolumn \else \newpage \fi
        \def\thefootnote{\arabic{footnote}}
        \setcounter{footnote}{0}}  

\catcode`@=12
\relax

\makeatletter
\newcounter{pubctr}
\def\publist{\@ifnextchar[{\@publist}{\@@publist}}
\def\@publist[#1]{\list
        {[\arabic{pubctr}]\hfill}{\settowidth\labelwidth{[999]}
        \leftmargin\labelwidth
        \advance\leftmargin\labelsep
        \@nmbrlisttrue\def\@listctr{pubctr}
        \setcounter{pubctr}{#1}\addtocounter{pubctr}{-1}}}
\def\@@publist{\list
        {[\arabic{pubctr}]\hfill}{\settowidth\labelwidth{[999]}
        \leftmargin\labelwidth
        \advance\leftmargin\labelsep
        \@nmbrlisttrue\def\@listctr{pubctr}}}
 \relax
\makeatother
\newskip\humongous \humongous=0pt plus 1000pt minus 1000pt

\newif\ifdtup

\relax

\def\d{\partial}

\def\sqr#1#2{{\vcenter{\vbox{\hrule height.#2pt\hbox{\vrule width.#2pt 
height#1pt \kern#1pt \vrule width.#2pt}\hrule height.#2pt}}}}

\def\=d{\,{\buildrel\rm def\over =}\,}

\def\i3p{\p32\int d^3p}

\def\As{A\hbox to 1pt{\hss /}}
\def\np4{\int d^4p_1\cdots d^4p_{n-1}\, }

\def\sgn{{\rm sgn}\, }

\def\nx4{\int d^4x_1\ldots d^4x_n\, }

\def\kon#1#2{\vbox{\halign{##&&##\cr
\lower4pt\hbox{$\scriptscriptstyle\vert$}\hrulefill &
\hrulefill\lower4pt\hbox{$\scriptscriptstyle\vert$}\cr $#1$&
$#2$\cr}}}

\def\konv#1#2#3{\hbox{\vrule height12pt depth-1pt}
\vbox{\hrule height12pt width#1cm depth-11.6pt}
\hbox{\vrule height6.5pt depth-0.5pt}
\vbox{\hrule height11pt width#2cm depth-10.6pt\kern5pt
      \hrule height6.5pt width#2cm depth-6.1pt}
\hbox{\vrule height12pt depth-1pt}
\vbox{\hrule height6.5pt width#3cm depth-6.1pt}
\hbox{\vrule height6.5pt depth-0.5pt}}
\def\konu#1#2#3{\hbox{\vrule height12pt depth-1pt}
\vbox{\hrule height1pt width#1cm depth-0.6pt}
\hbox{\vrule height12pt depth-6.5pt}
\vbox{\hrule height6pt width#2cm depth-5.6pt\kern5pt
      \hrule height1pt width#2cm depth-0.6pt}
\hbox{\vrule height12pt depth-6.5pt}
\vbox{\hrule height1pt width#3cm depth-0.6pt}
\hbox{\vrule height12pt depth-1pt}}

\def\konw#1#2#3{\hbox{\vrule height12pt depth-1pt}
\vbox{\hrule height12pt width#1cm depth-11.6pt}
\hbox{\vrule height6.5pt depth-0.5pt}
\vbox{\hrule height12pt width#2cm depth-11.6pt \kern5pt
      \hrule height6.5pt width#2cm depth-6.1pt}
\hbox{\vrule height6.5pt depth-0.5pt}
\vbox{\hrule height12pt width#3cm depth-11.6pt}
\hbox{\vrule height12pt depth-1pt}}

\def\i{{\rm int}}

\def\m3{{\mu_1\mu_2\mu_3}}

\def\p{{(+)}}

\def\be{\begin{equation}}       \def\eq{\begin{equation}}
\def\ee{\end{equation}}         \def\eqe{\end{equation}}

\def\bea{\begin{eqnarray}}      \def\eqa{\begin{eqnarray}}
\def\ena{\end{eqnarray}}        \def\eea{\end{eqnarray}}
                                \def\eqae{\end{eqnarray}}

\def\ba{\begin{array}}
\def\ea{\end{array}}
\def\unit{1 \hskip-.3em \raise2pt\hbox{$ \scriptstyle |$ } }

\def\d{\delta}
\def\f{\phi}               

\def\i{\iota}

\def\m{\mu}

\def\p{\pi}                
\def\t{\tau}

\def\cd{{\cal D}}

\def\cj{{\cal J}}
\def\ck{{\cal K}}
\def\cl{{\cal L}}

\def\bop#1{\setbox0=\hbox{$#1M$}\mkern1.5mu
        \vbox{\hrule height0pt depth.04\ht0
        \hbox{\vrule width.04\ht0 height.9\ht0 \kern.9\ht0
        \vrule width.04\ht0}\hrule height.04\ht0}\mkern1.5mu}
\def\Box{{\mathpalette\bop{}}}                        
\def\pa{\partial}                              

\def\>{\rangle} 

\def\<{\langle} 
\def\Dsl{D \hskip-.6em \raise1pt\hbox{$ / $ } }

\def\sl#1{\rlap{\hbox{$\mskip 1 mu /$}}#1}
\def\leftrightarrowfill{$\mathsurround=0pt \mathord\leftarrow \mkern-6mu
       \cleaders\hbox{$\mkern-2mu \mathord- \mkern-2mu$}\hfill
       \mkern-6mu \mathord\rightarrow$}
\def\dvec#1{\vbox{\ialign{##\crcr
       \leftrightarrowfill\crcr\noalign{\kern-1pt\nointerlineskip}
       $\hfil\displaystyle{#1}\hfil$\crcr}}}          
\def\hook#1{{\vrule height#1pt width0.4pt depth0pt}}
\def\leftrighthookfill#1{$\mathsurround=0pt \mathord\hook#1
       \hrulefill\mathord\hook#1$}
\def\underhook#1{\vtop{\ialign{##\crcr                 
       $\hfil\displaystyle{#1}\hfil$\crcr
       \noalign{\kern-1pt\nointerlineskip\vskip2pt}
       \leftrighthookfill5\crcr}}}
\def\smallunderhook#1{\vtop{\ialign{##\crcr      
       $\hfil\scriptstyle{#1}\hfil$\crcr
       \noalign{\kern-1pt\nointerlineskip\vskip2pt}
       \leftrighthookfill3\crcr}}}

\def\sfrac#1#2{{\vphantom1\smash{\lower.5ex\hbox{\small$#1$}}\over
       \vphantom1\smash{\raise.4ex\hbox{\small$#2$}}}} 
\def\bfrac#1#2{{\vphantom1\smash{\lower.5ex\hbox{$#1$}}\over
       \vphantom1\smash{\raise.3ex\hbox{$#2$}}}}      
\def\afrac#1#2{{\vphantom1\smash{\lower.5ex\hbox{$#1$}}\over#2}}  
\def\on#1#2{{\buildrel{\mkern2.5mu#1\mkern-2.5mu}\over{#2}}}
\def\ddt#1{\on{\hbox{\LARGE .\kern-2pt.}}#1}             
\def\tdt#1{\on{\hbox{\LARGE .\kern-2pt.\kern-2pt.}}#1}   

\def\boxes#1{
       \newcount\num
       \num=1
       \newdimen\downsy
       \downsy=-1.5ex
       \mskip-2.8mu
       \bo
       \loop
       \ifnum\num<#1
       \llap{\raise\num\downsy\hbox{$\bo$}}
       \advance\num by1
       \repeat}
\def\boxup#1#2{\newcount\numup
       \numup=#1
       \advance\numup by-1
       \newdimen\upsy
       \upsy=.75ex
       \mskip2.8mu
       \raise\numup\upsy\hbox{$#2$}}

\newskip\humongous \humongous=0pt plus 1000pt minus 1000pt

\newif\ifdtup

\def\to{\rightarrow}

\def\1ov4{{1\over 4}}

\def\pa{\partial}

\def\ddt{\dot{\t}}

\def\pa{\partial}

\def\nonu{\nonumber \\{}}

\begin{document}



\thispagestyle{empty} \setcounter{footnote}{0}
\begin{flushright}
MPI/PhT-98-85\\
SPIN-1998/6\\ 
hep-th/9811231
\end{flushright}

\vskip 1,7cm
\begin{center}
\begin{center}
{\large\bf The Quantum Noether Condition \\
in terms of Interacting Fields}
\end{center}
\vskip 2.0cm
\centerline{\bf Tobias Hurth\footnote{ hurth@mppmu.mpg.de}}
\vskip 0.5cm
\centerline{\it Max-Planck-Institute for Physics, Werner-
Heisenberg-Institute}
\centerline{\it F\"ohringer Ring 6, D-80805 Munich, Germany}
  \vskip 0,5cm
  {\bf Kostas Skenderis 
\footnote{K.Skenderis@phys.uu.nl}}
  \vskip 0,2cm
  {\it Spinoza Institute, University of Utrecht,\\ 
 Leuvenlaan 4, 3584 CE Utrecht, The Netherlands}
\end{center}
\vskip 1,0cm
{\bf Abstract.} -
{\small
We review our recent work, hep-th/9803030, on 
the constraints imposed by global or local 
symmetries on perturbative quantum field theories.
The analysis is performed in the Bogoliubov-Shirkov-Epstein-Glaser 
formulation of perturbative quantum field theory. In this formulation 
the S-matrix is constructed directly in the asymptotic Fock space
with only input causality and Poincar\'{e} invariance.
We reformulate the symmetry condition proposed in our earlier
work in terms of interacting Noether 
currents.
\vskip 3,0cm
\begin{center}
In {\it New Developments in Quantum Field Theory},\\
Springer, eds. P. Breitenlohner, D. Maison and J. Wess
\end{center}
}

\newpage

\section{Introduction}
\setcounter{equation}{0}

The relation between symmetries and quantum theory is 
an important and fundamental issue. For instance,
symmetry relations among correlation functions (Ward identities) 
are often used in order to prove that a quantum field theory 
is unitary and renormalizable.
Conversely, the  violation of a classical symmetry at the quantum 
level (anomalies) often indicates that the theory is inconsistent.
Furthermore, in recent years symmetries (such as supersymmetry)
have been instrumental in uncovering non-perturbative 
aspects of quantum theories (see, for example, \cite{SW}).
It is, thus, desirable to understand the interplay 
between symmetries and quantization in a manner which is free of the 
technicalities inherent in the conventional Lagrangian 
approach (regularization/renormalization) and in  
a way which is model independent as much as possible.

In a recent paper\cite{paper1} we have presented 
a general method, the Quantum Noether Method, for 
constructing perturbative quantum field theories 
with global symmetries. Gauge theories are 
within this class of theories, the global symmetry being 
the BRST symmetry\cite{BRST}. The method is 
established in the causal approach 
to quantum field theory introduced by Bogoliubov and Shirkov \cite{BS}
and developed by Epstein and Glaser\cite{EG0,stora}.
This explicit construction method rests
directly on the axioms of relativistic quantum field theory. 
The infinities encountered in the conventional approach 
are avoided by a proper handling of the correlation 
functions as operator-valued distributions. 
In particular, the well-known problem of ultraviolet (UV)
divergences is reduced to the mathematically 
well-defined problem of splitting an operator-valued distribution with causal
support into a distribution with retarded and a distribution with
advanced support or, alternatively \cite{stora, fredenhagen1}, 
to the continuation of time-ordered products to coincident points.
Implicitly, every consistent renormalization scheme
solves this problem. Thus, the explicit Epstein-Glaser (EG)
construction should not be regarded as a special renormalization 
scheme but as a general framework in which the conditions posed by the 
fundamental axioms of quantum field theory
(QFT) on any renormalization scheme are built in by
construction. In this sense our method is  
independent from the causal framework. Any  
renormalization scheme can be used to work out the
consequences of the general symmetry conditions proposed in \cite{paper1}.

In the EG approach the $S$-matrix is directly constructed
in the Fock space of free asymptotic fields in a form of formal
power series. The coupling constant is replaced by a tempered
test function $g(x)$ (i.e. a smooth function rapidly decreasing 
at infinity) which switches on the interaction. Instead
of evaluating the $S$-matrix by first computing 
off-shell Greens functions by means of Feynman rules
and then applying the LSZ formalism, the $S$-matrix is 
directly obtained by imposing causality and Poincar\'{e}
invariance. The method can be regarded as an ``inverse''
of the cutting rules. One builds $n$-point functions out 
of $m$-point functions ($m<n$) by suitably 
``gluing'' them together. The precise manner 
in which this is done
is dictated by causality and Poincar\'{e} invariance  
(see appendix A for details). 
One shows, that this process uniquely fixes the $S$-matrix 
up to local terms (which we shall call ``local normalization terms''). 
At tree level these local terms are nothing but 
the Lagrangian of the conventional approach\cite{paper1}. 

The problem we set out to solve in \cite{paper1} was to 
determine how to obtain a quantum theory 
which, on top of being causal and Poincar\'{e} invariant,
is also invariant under a global symmetry.
For linear symmetries such as global internal symmetries
or discrete $C$, $P$, $T$ symmetries the solution 
is well-known: one implements the symmetry in the 
asymptotic Fock space space by means of an (anti-) unitary 
transformation. 
The focus of our investigation in \cite{paper1} was 
symmetries that are 
non-linear in the Lagrangian formulation. The prime 
examples are BRST symmetry and supersymmetry (in the 
absence of auxiliary fields). The main puzzle is 
how a theory formulated in terms of asymptotic 
fields only knows about the inherent non-linear structure.

The solution to the problem is rather natural. One imposes
that the Noether current that generates 
the asymptotic symmetry is conserved 
at the quantum level, i.e. inside correlation functions.
This condition, the Quantum Noether Condition (QNC),
constrains the local normalization terms left 
unspecified by causality and Poincar\'{e} invariance.
At tree-level one finds that the asymptotic Noether 
current renormalizes such that it generates 
the full non-linear transformation rules.
At the quantum level the same condition yields
the corresponding Ward identities. 
The way the methods works is analogous to the classical
Noether method \cite{deser,sugra}, hence its name.
In addition, we have shown that the QNC
is equivalent to the condition that the 
$S$-matrix is invariant under the symmetry under 
question (i.e. the $S$-matrix commutes with the 
generator of the asymptotic symmetry).

Quantum field theory, however, is usually formulated in terms of interacting 
fields. In the Lagrangian formulation, the symmetries of the theory 
are the symmetries of the action (or more generally of the 
field equations) that survive at the quantum level. 
These symmetries are generated by interacting Noether currents.
It will, thus, be desirable to express the QNC in terms 
of the latter. As we shall see, this is indeed possible.
The QNC in term of the interacting current is given 
in (\ref{cond3}).
If the symmetry is linear then the condition 
is that the interacting current is conserved (as expected).
If the symmetry, however, is non-linear the interacting 
current is only conserved in the adiabatic limit ($g \to$ const.).

One important example is Yang-Mills theory. In this 
case, the corresponding Noether current is the BRST 
current. 
Because there are unphysical degrees of freedom present in gauge theories,
one needs a subsidiary condition in order  
to project out the unphysical states. 
The subsidiary condition should remain invariant under time 
evolution. This means that it should be expressed in terms of a conserved 
charge. The appropriate charge for gauge theories is the 
BRST charge \cite{KO}. The subsidiary condition is that 
physical states should be annihilated by the BRST charge $Q_{int}$
(and not be $Q_{int}$-exact). 

The considerations in \cite{KO}, however, (implicitly) assumed the
naive adiabatic limit. For pure gauge theories this limit 
seem not to exist. Then from the Quantum Noether Condition (\ref{cond3})
follows that the interacting BRST current
is not conserved before the adiabatic limit.
We stress, however, that the Quantum Noether Condition 
allows one to work out all consequences of non-linear symmetries for 
time-ordered operator products before the adiabatic limit is taken. 
As we shall see, one can even identify the non-linear
transformation rules.   

We organize this paper as follows:
In the next section we shortly review the Quantum Noether Method.
In section 3 we express the Quantum Noether Condition 
in terms of the interacting Noether current.
Section 4 contains a discussion of future 
directions. In the appendix we present 
the main formulae of the causal framework 
and our conventions.

\section{The Quantum Noether Method}
\setcounter{equation}{0}

In the EG approach one starts with a set of free  fields in the 
asymptotic Fock space. These fields satisfy their (free) field
equations and certain commutation relations.  To define the theory
one still needs to specify $T_1$, the first term in the $S$-matrix.  
(Actually, as we shall see, even $T_1$ is not free in our 
construction method but is also constrained  by the Quantum Noether Condition).
Given $T_1$ one can, in a well defined manner, construct iteratively the 
perturbative $S$-matrix. The requirements of causality and Poincar\'{e} 
invariance completely fix the $S$-matrix up to local terms.
The additional requirement that the theory is invariant under 
a global and/or local symmetry imposes constraints on these local terms.

To construct  a theory with global and/or local
symmetry we introduce the coupling $g_\m j^\m_0$ in the theory,
where $j^\m_0$ is the Noether current that generates the 
asymptotic (linear) symmetry transformations, and  
we impose the condition that ``the Noether current 
is conserved at the quantum level'' 
\be \label{cons}
\pa_\m \cj_n^{\m} (x_1, \cdots, x_n; \hbar) =0,
\ee
where we introduce the notation (we
use the  abbreviation $\pa/ \pa x^\m_l = \pa^l_\m$)
\be \label{nota}
\pa_\m \cj_n^{\m} (x_1, \cdots, x_n; \hbar)= 
\sum_{l=1}^n \pa_\m^l \cj_{n/l}^{\m}, 
\ee
and 
\be
\cj^\m_{n/l}=T[T_1 (x_1) \cdots j_0^\m(x_l) \cdots T_1(x_n)].
\ee 
(for $n=1$, $\cj^\m_1(x_1)=j_0^\m(x_1)$).
In other words we consider an $n$-point function
with one insertion of the current $j_0^\m$ at the point $x_l$.
Notice that since the left hand side of (\ref{cons}) is a formal
Laurent series in $\hbar$, this condition is actually a set of conditions. 

One may apply the inductive EG construction to work out the consequences of 
(\ref{cons}). This may be done by first working out $T[j_0 T_1...T_1]$
and then constructing (\ref{nota}). However, there is an alternative
route  \cite{paper1}. One relaxes 
the field equations of the fields $\phi^A$. Then the 
inductive hypothesis takes the form: for $m<n$,
\be \label{tfeq}
\sum_{l=1}^m \pa^l_\mu \cj_{m/l}^{\m} =
\sum_{A} R^{A;m}(\hbar) \ck_{AB} \f^B \d(x_1, \ldots, x_m), 
\ee 
where 
\be \label{feq}
\ck_{AB} \f^B= \pa^\m {\pa \cl_0 \over \pa (\pa^\m \f^A)} 
- {\pa \cl_0 \over \pa \f^A}   
\ee
are the free field equations ($\cl_0$ is the free Lagrangian 
that yields (\ref{feq}); the present formulation assumes 
that such a Lagrangian exists).
The coefficients $R^{A;m}(\hbar)$ are defined by (\ref{tfeq})
and are formal series in $\hbar$.

Clearly, if we impose the field equation we go back to (\ref{cons}).
The converse is also true. Once one relaxes the field 
equations in the inductive step,  (\ref{cons}) implies (\ref{tfeq})
as was shown in \cite{paper1}. The advantage of the off-shell 
formulation is that it makes manifest 
the non-linear structure: the coefficients $R^{A;m}(\hbar)$
are just the order $m$ part 
of the non-linear transformation rules. In addition, 
the calculation of local on-shell terms arising from tree-level graphs
simplifies:

We now discuss the condition (\ref{cons}) 
at tree-level. For the analysis at loop level we refer to \cite{paper1}.
At tree-level we only need the $\hbar^0$ part of (\ref{tfeq}).
Let us define 
\be
\label{delta}
s_{(m-1)}\f^A = {1 \over  m!} R^{A;m}(\hbar^0).
\ee 
Depending on the theory under consideration the quantities $R^{A;m}(\hbar^0)$
may be zero after some value of $m$. Without loss of generality we 
assume that they are zero for $m>k+1$, for some integer $k$ (which 
may be infinity; the same applies for $k'$ below.).
One shows that 
\be \label{fulltr}
s \f^A = \sum_{m=0}^k g^m s_m \f^A
\ee
are symmetry transformation rules that leave the Lagrangian, 
\be \label{lagr}
\cl = \sum_{m=0}^{k'} g^m \cl_m,
\ee 
invariant (up to total derivatives),
where $k'$ is also an integer (generically not equal to $k$).
The Lagrangian $\cl$ will be determined from the tree-level normalization
conditions as follows,
\be \label{lagdef} 
\cl_m = {\hbar \over i} {N_m \over m!}, \quad {\rm for} \quad m>1,
\ee
where $N_m$ denotes the local normalization ambiguity of  
$T_m[T_1(x_1)...T_1(x_m)]$ in tree graphs defined with respect 
to the naturally split solution (i.e. the Feynman propagator
is used in tree-graphs). For $m=1$, $\cl_1=(\hbar/i)T_1$.
The factor $m!$ reflects the fact that 
$T_m[...]$ appears in (\ref{GC}) with a combinatorial 
factors $m!$ while the factor $\hbar/i$ is there to cancel the 
overall factor $i/\hbar$ that multiplies the action in the 
tree-level $S$-matrix. Notice that we regard 
(\ref{lagdef}) as definition of $\cl_m$. 
Let us further define $j_n^\m$ as the local normalization 
ambiguity of $T_n[j_0T_1...T_1]$,\footnote{
We use the following abbreviations for the delta function distributions
$\d^{(n)}=\d(x_1, \ldots, x_n)=$\\ $\d(x_1-x_2)\cdots\d(x_{n-1}-x_n)$.}
\be \label{jndef}
T_n [j_0^\m(x_1) T_1(x_2) \cdots T_1 (x_n)]=
T_{c,n} [j_0^\m(x_1) T_1(x_2) \cdots T_1(x_n)] 
+ j_{n-1}^\m \d^{(n)}
\ee
where $T_{n,c}$ denotes the naturally splitted solution.
We shall see that the normalization terms $j_n$ 
complete the asymptotic current $j_0$ to the 
Noether current that generates the non-linear symmetry 
transformations (\ref{fulltr}).

We wish to calculate the tree-level terms at $n$th order.
The causal distribution $\sum_{l=1}^n \pa_\m^l \cd^\m_{n/l}$ 
at the $n$th order consists of a sum of terms each of these being  
a tensor product of 
$T_m[T_1 ... T_1 \partial{\cdot}j_0 T_1 ... T_1)$ ($m<n$) with 
$T$-products that involve 
only $T_1$ vertices according to the general 
formulae (\ref{ret},\ref{adva},\ref{D-dist}).
By the off-shell induction hypothesis, we have for all $m<n$ 
\be \label{offshell}
\sum_{l=1}^m \pa^l_\mu \cj_{m/l}^{\m} =
\sum_{A} (m! s_{m-1} \f^A)  \ck_{AB} \f^B \d^{(m)}. 
\ee
As explained in detail in \cite{paper1}, at order $n$ one 
obtains all local on-shell terms by performing the so-called 
``relevant contractions'', namely the
contractions between the $\f^B$ in the right hand side of (\ref{offshell})
and $\f$ in local terms.
In this manner we get the following general formula for the 
local term $A_{c,n}$  arising through tree-level contractions at level
$n$,
\be \label{loc}
A_{c,n}(tree) = \sum_{\pi \in \Pi^n} \sum_{m=1}^{n-1}
\pa_\m \cj_m^\m(x_{\p(1)}, \ldots, x_{\p(m)}) 
N_{n-m}\d(x_{\p(k+1)}, \ldots, x_{\p(n)})
\ee
where it is understood that in the right hand side only 
``relevant contractions'' are  
made. The factors $N_{n-m}$ are tree-level normalization terms of 
the $T$-products that contain $n-m$ $T_1$ vertices. 

In \cite{paper1} we have provided a detailed analysis 
of (\ref{loc}) for any $n$ (under the assumption that 
the Quantum Noether Method is not obstructed). In the next section, 
we will need these results in order to show that condition 
(\ref{cond3}) is equivalent to condition (\ref{cons}).
We therefore list them here without proofs.

The $n=1$ case is trivial. One just gets that $R^{A;1}(\hbar^0)=s_0 \f^A$. 
For $2 \leq n \leq k+1$, the 
condition (\ref{tfeq}) at tree-level yields the following constraint
on the local normalization terms of the $T_m$, $m<n$,
\be \label{n<k}
s_0 \cl_{n-1} + s_1 \cl_{n-2} + \cdots + s_{n-2} \cl_1=
\pa_\mu \cl^\m_{n-1} + s_{n-1}\f^A \ck_{AB}\f^B
\ee
and, furthermore, determines $j_{n-1}^\m$,
\be \label{jn}
j_{n-1}^\m= -n!\cl_{n-1}^\m
+(n-1)! \sum_{l=0}^{n-2} (l+1) \frac{\pa \cl_{n-1-l}}
{\pa(\pa_\m \f^A)} s_l \f^A. 
\ee

For $n>k+1$ we obtain,
\be \label{n>k}
s_0 \cl_{n-1} + s_1 \cl_{n-2} 
+ \cdots + s_k \cl_{n-1-k}=
\pa_\m \cl_{n-1}^\m,
\ee
and 
\be \label{jn1}
j_{n-1}^\m=-n!\cl^\m_{n-1} + (n-1)! \sum_{l=1}^{k} 
l \frac{\pa \cl_{n-l}}{\pa(\pa_\m \f^A)} s_{l-1} \f^A.
\ee
Depending on the theory under consideration the $\cl_n$'s will 
be zero for $n>k'$, for some integer $k'$. Given the integers $k$ and
$k'$, there is also an integer $k''$ (determined from the other two)
such that $\cl^\m_n=0$, for $n>k''$. 

Summing up the necessary and sufficient conditions (\ref{n<k}), 
(\ref{n>k}) for the Quantum Noether method to 
hold at tree level we obtain,
\be
s \sum_{l=1}^{k'} g^l \cl_l = \sum_{l=1}^{k''} \pa_\m \cl_l^\m 
+ (\sum_{l=1}^k g^l s_l \f^A) \ck_{AB} \f^B
\ee
Using $s_0 \cl_0 = \pa_\m k^\m_0$ and for $l \leq k$
\be
s_l \f^A \ck_{AB} \f^B = \pa_\m ({\pa \cl_0 \over \pa(\pa_\m \f^A)} s_l \f^A)
-s_l \cl_0
\ee
we obtain,
\be \label{treecon}
s \cl = \pa_\m (\sum_{l=0}^{k''} g^l k_l^\m)
\ee 
where, for $1<l \leq k$, 
\be \label{kdef}
k_l^\m = \cl_l^\m + {\pa \cl_0 \over \pa(\pa_\m \f^A)} s_l \f^A
\ee
and for $l>k$, $k_l^\m = \cl_l^\m$.
We therefore find that $\cl$ is invariant under the symmetry 
transformation,
\be
s \f^A = \sum_{l=0}^k g^l s_l \f^A.
\ee
According to Noether's theorem there is an associated  Noether current.
One may check that the current normalization terms $j_m^\m$ 
(\ref{jn}), (\ref{jn1}) are in one-to-one correspondence
with the terms in the Noether current.
Therefore the current $j_0$ indeed renormalizes to the 
full non-linear current.

\section{Conservation of the Interacting Noether Current}
\setcounter{equation}{0}

The Quantum Noether Condition (\ref{cons}) can be reformulated in terms of 
interacting fields.
Let $j^\m_{0,int}$ and $\tilde{j}^\m_{1,int}$ 
be the interacting currents corresponding to  free field operators 
$j_0^\m$ and  $\tilde{j}_1^\m$, 
respectively, perturbatively constructed 
according to (\ref{defint}). $\tilde{j}_1^\m$ is equal to   
$- \cl^\m_1$ (defined in (\ref{n<k})) as will see below.
Then the general Ward identity 
\be \label{cond3}
\pa_\m j^\m_{0,int} = \pa_\mu g \tilde{j}^\mu_{1, int} 
\ee
is equivalent to condition (\ref{cons}).
According to condition (\ref{cond3}) 
the interacting Noether current $j^\m_{0,int}$ is conserved
only if it generates a linear symmetry, i.e. $\tilde{j}_1^\m$ vanishes,
or otherwise in the adiabatic limit  
\mbox{$g(x)\rightarrow 1$}, provided this limit exists.
In the following we shall show that the condition (\ref{cond3})
yields the same conditions on the 
the time-ordered products $T_n [T_1... T_1]$
as the Quantum Noether condition (\ref{cons}).
In this sense the two general symmetry conditions
are considered equivalent.

Because Poincar\'{e} invariance and causality already 
fix the time-ordered products $T_n [T_1... T_1]$
up to the local normalization ambiguity $N_n$, we only have to 
show that these local normalization terms  $N_n$ are constrained 
in the same way by both conditions, (\ref{cond3}) and (\ref{cons}). 

First, we translate the condition (\ref{cond3}) to a condition on
time-ordered products using the formulae given in the appendix: 

The perturbation series for the interacting field operator  $j^\mu_{int}$
of a free field operator $j^\mu$ is given by 
the advanced distributions of the corresponding expansion of 
the $S$-matrix (see (\ref{defint})):
\be
\label{advanced}
j_{int}^\mu (g,x) =  j^\mu (x) + \sum_{n=1}^\infty \frac{1}{n!} 
\int d^4 x_1 \ldots d^4 x_n  
Ad_{n+1} \left[T_1 (x_1) \ldots T_1 (x_n); 
j^\mu (x) \right] g(x_1) \ldots g(x_n),
\ee
where $Ad_{n+1}$ denotes the advanced operator-valued 
distribution with 
$n$ vertices $ T_1 $ and one vertex $ j^\mu (x) $ at the 
$(n+1)$th position. 
This distribution is only symmetric in the first $n$  
variables $x_1, \ldots, x_n $. 
The support properties are defined with respect to the 
unsymmetrized variable $x$.

With the help of (\ref{advanced}), we rewrite the left hand side of 
equation (\ref{cond3})
\be
\partial_\mu^x j_{0,int}^{\mu} (x) = \pa_\mu^x j_0^\m(x) +
\sum_{n=1}^\infty \frac{1}{n!} 
\int d^4 x_1 \ldots d^4 x_n 
\partial_\mu^x Ad_{n+1} \left[ T_1 (x_1) 
\ldots T_1 (x_n); j_0^\mu (x) \right] 
g(x_1) \ldots g(x_n)  
\ee
and the right hand side of (\ref{cond3})
\bea
\tilde{j}_{1,int}^{\mu} (x) \partial_\mu g(x) & = & 
\sum_{n=0}^\infty \frac{1}{n!} 
\int d^4 x_1 \ldots d^4 x_n d^4 x_{n+1} \\ 
& & Ad_{n+1} \left[ T_1 (x_1) \ldots T_1 (x_n); 
\tilde{j}_1^\mu (x) \right] 
\delta (x-x_{n+1}) \quad g(x_1) \ldots g(x_n)
\partial_\mu^{x_{n+1}} g(x_{n+1}) \nonumber 
\eea
After partial integration, symmetrization of the 
integrand in the variable 
$(x_1, \ldots, x_{n+1})$ and 
shifting the summation index, the right hand side of (\ref{cond3}) 
can be further rewritten as
\bea
\tilde{j}_{1,int}^{\mu} (x) \partial_\mu 
g(x) & = & - \sum_{n=1}^\infty \frac{1}{n!} 
\int d^4 x_1 \ldots d^4 x_n \\
&&\sum_{j=1}^{n} \left\{ Ad_{n} 
\left[T_1(x_1) \ldots \widehat{T(x_j)} \ldots T(x_n);\tilde{j}_1^\mu(x) 
\right] 
\partial_\mu^{x_j} \delta (x_j - x) \right\} g(x_1) \ldots g(x_n) \nonumber
\eea
where the hat  indicates that this coupling has to be omitted.  
Equation (\ref{cond3}) reads then 
\bea
\label{cond3adv}
&& \pa_\mu j^\m_0 =0, \qquad (n=0)\nonu
&& \partial_\mu^x Ad_{n+1} \left[ T_1(x_1) 
\ldots T(x_n); j_0^\mu (x) \right] \nonu  
&&+ \sum_{j=1}^{n} Ad_n \left[ T_1(x_1) \ldots \widehat{T(x_j)} 
\ldots T(x_n); \tilde{j}_1^\mu (x) \right]\partial_\mu^{x_j} 
\delta (x_j - x) = 0,  \quad (n>0)
\eea
where the local normalization terms of the $ Ad $-distributions with
respect to a specified splitting solution will be given  below.

In the following we discuss the equivalent condition of the 
time-ordered distributions 
instead of the advanced ones in order to compare the unsymmetrized 
condition (\ref{cond3})
with the symmetrized Quantum Noether Condition (\ref{cons}). 
We get  
instead of (\ref{cond3adv})
\bea
&& \partial_\mu^x T_{n+1} \Big[ T_1(x_1) \ldots T_1(x_n); j_0^\mu (x) \Big] 
\nonumber \\ 
&&\hspace{2cm} = -\sum_{j=1}^{n} T_n \left[ T_1(x_1) \ldots 
\widehat{T_1(x_j)} \ldots T_1(x_n); 
\tilde{j}_1^\mu (x) \right] \partial_\mu^{x_j} 
\delta (x_j - x) \label{condt}
\eea
These distributions get smeared out by $ {g(x_1) 
\ldots g(x_n) \tilde {g}(x)}$, where 
the test-function $\tilde{g}$ differs from $g$. 
One easily verifies the left hand side of (\ref{condt}) is 
just the Quantum Noether Condition (\ref{cons}) but without the 
symmetrization; the missing symmetrization produces 
the extra terms on the right hand side of (\ref{condt}) as we shall see.

We shall use the same off-shell procedure in order to fix the 
local on-shell obstruction 
terms (which is explained in detail in \cite{paper1}, section 4.2).
The starting point ($n=0$) of both conditions  is the same  
\be
\partial_\mu j_0^\mu (x) = s_0 \phi^A \ck_{AB} \phi^B 
\ee
We have now for $n=1$,
\be
\partial_\mu^x \left( T_{2,c} 
[T_1 (x_1) j_0^\mu (x)] + j_1^\mu  \delta (x_1 - x) \right)
=  - \tilde{j}_1^\mu (x) \partial_\mu^{x_1} \delta (x_1 - x)
\ee
Working out the left hand side (and using  
$T_1= \frac{i}{\hbar} \cl_1$) we obtain,
\be \label{n=2new}
\partial_\mu^x \left( j_1^\mu \delta (x_1 - x) \right) + 
s_0 \cl_1 \delta (x_1 - x) - 
\partial_\mu^x \left( \frac {\partial \cl_1} 
{\partial (\partial_\mu \phi^A)} s_0 \f^A \delta (x_1 - x) \right) 
= \tilde{j}_1^\mu (x) \partial_\mu^{x} \delta (x_1 - x) 
\ee
This condition fixes the local renormalization 
of $ j_0^\mu $ at order $g$, denoted by  $ j_1^\mu $
(defined with respect to the natural splitting solution $T_{2,c}$) 
and also $ \tilde{j}_1^\mu $ in condition (\ref{cond3}). 
The latter term, proportional  
to the derivative of the $\delta$-distribution, is left 
over in our new unsymmetrized 
condition. Note that in the symmetrized case, 
we reduced these 
kind of terms to ones  proportional 
to the $\delta$-distribution with the help of 
distributional identities. 

The condition (\ref{n=2new}) can be fulfilled for some 
local operators $ j_1^\mu $ and $ \tilde{j}_1^\mu $ if and only if 
$s_0 \cl_1 $ is a divergence up to field equation terms,
\be
s_0 \cl_1 = \partial_\mu \cl_1^\mu + s_1 \phi^A \ck_{AB} \phi^B.
\ee
In the absence of real obstructions 
this equation has solutions and we get
\be \label{j1}
j_1^\mu = - \cl_1^\mu + \frac {\partial \cl_1} 
{\partial (\partial_\mu \phi^A)} s_0 \phi^A
\ee
as local renormalization of $ j_{0,int}^\mu $ at order $ g^1 $ and 
\be
\tilde{j}_1^\mu = - \cl_1^\mu. 
\ee
Equation (\ref{j1}) should be compared with the analogous formulae (\ref{jn})
for $n=2$ \footnote{Notice that $n$ in the present section
should be compared with $n+1$ in section 2.}. We finally have
\bea
\label{n=2newfinal}
\partial_\mu^x T_{2} \left[ T_1 (x_1) j_0^\mu (x) \right] + 
\tilde{j}_1^\mu (x) \partial_\mu^{x_1} \delta (x_1 - x) = 
s_1 \phi^A \ck_{AB} \phi^B \d(x_1-x).  
\eea

The off-shell term on the right hand side of (\ref{n=2newfinal}) 
is responsible for local obstruction terms at the next order, $n=2$. 
We get (taking special care of 
derivative terms and advantage 
of our off-shell procedure): 
\bea
& \partial_\mu^x T_{3,c} \left[ T_1(x_1)T_1(x_2) j_0^\mu (x) \right] + 
\left( T_{2,c} \left[ T_1(x_1) \tilde {j}_1^\mu (x) \right] 
\partial_\mu^{x_2} \delta (x_2 - x) + [x_1 \leftrightarrow x_2] \right) & \\ 
& = {\hbar \over i} \left[
2s_1T_1 \delta^{(3)} - \left(2 \partial_\mu^x + \partial_\mu^{x_1} 
+ \partial_\mu^{x_2} 
\right) \left( \frac {\partial T_1} 
{\partial (\partial_\mu \phi^A)} s_1 \phi^A \delta^{(3)} \right) 
+ s_0 N_2 \delta^{(3)} - \partial_\mu^x \left( \frac {\partial N_2} 
{\partial (\partial_\mu \phi^A)} s_0 \phi^A \delta^{(3)} \right) \right]
 & \nonumber 
\label{n=222}
\eea
where $ N_2 $ denotes the tree-normalization term of 
$ T_2 [T_1 T_1] $ which is uniquely 
defined with  respect to the natural splitting solution  
$ T_{2,c} [T_1 T_1] $. Now we include also the 
normalization ambiguity of the 
other distributions involved: 
\bea
\label{norm22}
T_3 \left[ T_1(x_1) T_1(x_2) j_0^\mu (x) \right] &=& T_{3,c} 
\left[ T_1(x_1) T_1(x_2) j_0^\mu (x) \right] + 
j_2^\mu (x) \delta(x_1, x_2, x) \\ 
T_2 \left[ T_1 (x_i) \tilde{j}_1^\mu (x) \right] &=& T_{2,c} 
\left[ T_1 (x_i) \tilde{j}_1^\mu(x) \right] + 
\tilde{j}_{2}^\mu \delta (x_i - x) \nonumber 
\eea
According to  (\ref{n<k}) the Quantum Noether Condition (\ref{cons})
at order $n=3$ is fulfilled if and only if 
\bea
\label{n=2222}
s_1 \cl_1 + s_0 \cl_2 = \partial_\mu \cl_2^\mu + s_2 \phi^A \ck_{AB} \phi^B 
\eea
where the definition $ \cl_n = (\hbar/i) (N_n/n!) $ is used. 
Now the same is true for condition (\ref{n=222}). Only if 
(\ref{n=2222}) holds 
one can absorb the local terms  
on the right hand side of (\ref{n=222}) in the normalization terms 
$ j_2^\mu (x) $ and $\tilde{j}_2^\mu (x) $ 
given in (\ref{norm22}). 
The reasoning is again slightly different from the one in the  
symmetrized case. The distributions are only symmetric in the variables $x_i$, 
but $x$ is a distinguished variable. 
This means that the two local operator-valued distributions
\footnote{One could also choose as a basis 
$ \hat{A}_0^{'} \delta (x_1, x_2, x); \partial^x \left( \hat{A}_1^{'}  
\delta (x_1, x_2, x) \right) $.} 
\bea
\hat{A}_0  \delta (x_1, x_2, x); \quad \sum_{i=1}^2 \partial_{x_i} 
\left( \hat{A}_1 \delta (x_1, x_2, x) \right),\eea
where $ \hat{A}_0 (x) $ and $ \hat{A}_1 (x) $ are local operators, are 
independent (on the test functions 
$ \tilde{g} (x_1, x_2, x) := g (x_1) g (x_2) \tilde{g} (x)$ 
with $ g \neq  \tilde{g}$)\footnote{
In the symmetrized case, where one smears out with totally 
symmetric test functions $ g(x_1, x_2, x_3) := g (x_1) g(x_2) g(x_3)$, one has 
$\sum_{i = 1}^{2} \partial_{x_i} \left( \hat{A}_1 \delta(x_1, x_2, x) 
\right) = (2/3) \partial \hat{A}_1 
\delta (x_1, x_2, x).$}. 

So if and only if (\ref{n=2222}) is true the condition (\ref{cond3}) 
can be fulfilled at order 
$n=2$ and the local normalization terms of the interacting currents, 
$ j_{0,int}^\mu $ and $ \tilde{j}_{1,int}^\mu,$ get fixed to 
\bea
j_2^\mu &=& 2! \left( -  \cl_2^\mu + \frac {\partial \cl_2} {\partial 
(\partial_\mu \phi^A)} s_0 \phi^A + 
\frac {\partial \cl_1} {\partial (\partial_\mu \phi^A)} s_1 \phi^A \right) 
\nonu 
\tilde{j}_2^\mu &=& -2! \cl_2^\mu 
+ \frac{\pa \cl_1}{\pa (\pa_\m \f^A)} s_1 \f^A
\eea
Note the different symmetry factors in $ j_2^\mu $ compared with the 
symmetrized case (\ref{jn}). With these normalizations we get 
\bea
&\partial_\mu^x T_3 \left[ T_1 (x_1) T_1 (x_2) j_0^\mu (x) \right] + 
\left( T_2 \left[ T_1 (x_1) \tilde{j}_1 (x) \right] \partial_\mu^{x_2} 
\delta (x_2 - x) + [x_1 \leftrightarrow x_1] \right) & \\ 
& = 2! s_2 \phi^A \ck_{AB} \phi^B \delta(x_1, x_2, x) & \nonumber 
\eea
This corresponds to  (\ref{cond3}) at order $n=2$:
\bea
\partial_\mu^x j_{0,int}^{\mu} (x) \Big|_{g^2}=
\tilde{j}_{1,int}^{\mu} (x)\Big|_{g^1} \partial_\mu g (x)
+  2! s_2 \phi^A \ck_{AB} \phi^B (x).
\eea

{}From these first two steps of the inductive construction, one
already  realizes that 
in general the additional terms proportional to $\partial_\m  g$ 
in (\ref{cond3}) correspond to terms proportional to 
$\partial_\m \delta^n$ which are now independent. In the former condition 
(\ref{cons}) we got rid of these terms by  symmetrization and moding 
out the general formula $\sum_{l=1}^{n} \partial^l \delta^n=0$. This
formula is a direct consequence of translation invariance. 
Regardless this slight
technical difference both conditions, (\ref{cons}) and (\ref{cond3}),
pose the same consistency conditions on the physical normalization
ambiguity.

For $0 < n \leq k $, (where 
$k$ is the minimal integer such that $\forall m > k, s_m = 0$),
condition (\ref{condt}) yields
\bea 
&&\pa^x_\m(j_n \d^{(n+1)}) + 
n! \left( \sum_{l=0}^{n-1} s_{l} \cl_{n-l} \right) \d^{(n+1)}- \nonumber \\
&&-\sum_{l=0}^{n-1}\left
(n! \, \pa_\m^x + l \, (n-1)! \, \sum_{i=1}^n (\pa_\m^{x_i})
\right)
\left({\pa \cl_{n-l} \over \pa (\pa_\m \f^A)} s_{l} \f^A \d^{(n+1)} \right)= 
\tilde{j}^\m_n (x) \pa_\m^x \d^{(n+1)}
\eea
where $j_n^\m$ and $\tilde{j}^\m_n$ are defined by analogous 
to (\ref{norm22}) formulae. The sufficient and necessary condition 
for this equation to have a solution is 
\be 
s_0 \cl_n + \cdots + s_{n-1} \cl_1 = \pa_\m \cl^\m_n 
+ s_n \f^A \ck_{AB} \f^B.
\ee
This agrees with (\ref{n<k}) (we remind the reader 
that $n$ in present section  
corresponds to $n+1$ in section 2). Then the 
current normalization terms are given by
\bea \label{jn<k}
j^\mu_{n} &=& n! \left(- \cl_{n}^\mu + 
\sum_{l = 0}^{n-1} 
\frac{\partial \cl_{n-l}} {\partial (\partial_\mu \phi^A)} 
s_{l} \phi^A \right)\\ 
\tilde{j}^\mu_{n} &=& -n! \cl_{n}^\mu 
+ (n-1)! \sum_{l=0}^{n-1} l\,  
\frac{\partial \cl_{n-l}} {\partial (\partial_\mu \phi^A)} s_{l} \phi^A
\eea
and we have 
\bea
\label{final}
& &\partial_\mu^x j_{0,int}^{\mu} (x) \Big|_{g^{n}}
=\tilde {j}_{1,int}^{\mu} \Big|_{g^{n-1}} \partial_\mu g(x)
+ n! s_n \f^A \ck_{AB} \f^B (x)
\eea

For $n > k$, equation (\ref{condt}) yields
\bea 
&&\pa^x_\m(j_n \d^{(n+1)}) + 
n! \left( \sum_{l=0}^{k} s_{l} \cl_{n-l} \right) \d^{(n+1)} - \nonumber \\
&&-\sum_{l=0}^{k} \left(n! \pa_\m^x -l\, (n-1)! \sum_{i=1}^n (\pa_\m^{x_i})
\right)
\left({\pa \cl_{n-l} \over \pa (\pa_\m \f^A)} s_{l} \f^A \d^{(n+1)} \right)
= \tilde{j}^\m_n (x) \pa_\m^x \d^{(n+1)}
\eea
This equation now implies
\be
s_0 \cl_n + \cdots + s_k \cl_{n-k} = \pa_\m \cl^\m_n.
\ee
We further obtain for the current normalization terms,
\bea \label{jn>k}
j^\mu_{n} &=& n! \left(- \cl_{n}^\mu + 
\sum_{l = 0}^{k} 
\frac{\partial \cl_{n-l}} {\partial (\partial_\mu \phi^A)} 
s_{l} \phi^A \right)\\ 
\tilde{j}^\mu_{n} &=& -n! \cl_{n}^\mu 
+ (n-1)! \sum_{l=0}^{k} l\,  
\frac{\partial \cl_{n-l}} {\partial (\partial_\mu \phi^A)} s_{l} \phi^A
\eea
Therefore,
\bea
& &\partial_\mu^x j_{0,int}^{\mu} (x) \Big|_{g^{n}}
=\tilde {j}_{1,int}^{\mu} \Big|_{g^{n-1}} \partial_\mu g(x)
\eea
without using the free field equations.

In exactly the same way as in section 2, we deduce that 
the sum of all tree-level local normalization terms 
consitute a Lagrangian which is invariant (up to a 
total derivative) under the symmetry transformation
$s \f^A = \sum s_i \f^A$.  Inserting now the 
local normalization terms (\ref{jn<k}) and (\ref{jn>k})
into (\ref{advanced}) we obtain,
\be
j^\m_{0, int} = {\pa \cl \over \pa( \pa_\m \f^A)} s \f^A - k^\m
\ee
where we have used the definitions (\ref{lagr}), (\ref{fulltr}),
and (\ref{kdef}). The combinatorial factor $n!$ in (\ref{jn<k}) 
and (\ref{jn>k}) exactly cancels the same factor in (\ref{advanced}).
We, therefore, see that the interacting free current exactly 
becomes the full non-linear current.

We have, thus, found that going from 
condition (\ref{cons}) to condition (\ref{cond3}) just corresponds 
to a different technical treatment of the $\partial_\m \delta^{(n)}$ 
terms which
has no influence on the fact that both conditions pose the same 
conditions
on the normalization ambiguity of the physical $T_n$ distributions, 
namely
the consistency conditions of the classical Noether method.
Our analysis of the condition (\ref{cons}) at the loop level 
is also independent of this
slight technical rearrangement of the derivative terms. 
Thus, the issue of stability can be analyzed in exactly 
the same way as before (see section 4.3 of \cite{paper1}). One shows  
(under the assumption that the Wess-Zumino consistency 
condition has only trivial solutions) 
that condition (\ref{cond3}) at loop 
level also implies that the normalization ambiguity at the 
loop level, $N_n(\hbar)$,
is constrained in the same way as the tree-level normalizations, 
$N_n(\hbar^0)$. 
Once the stability has been established
the equivalence of (\ref{cons}) and (\ref{cond3}) at 
loop level follows.

Summing up, we have shown that conditions (\ref{cons})-(\ref{cond3})
yield all consequences of non-linear symmetries for time-ordered 
products before the adiabatic limit.
So at that level currents seem to be
sufficient. As mentioned in the introduction, however, 
if one wants to identify the physical Hilbert space,  
one may need 
to use the Noether charge $ Q_{int} = \int d^3x j^{0}_{int}(x) $.
As our Quantum Noether Condition (\ref{cond3}) shows, only in the 
adiabatic limit (provided the latter exists) the interacting 
Noether current is conserved. Moreover, there is an additional
technical obstacle. In the construction of the BRST charge a volume 
divergence occurs. In \cite{Fredenhagen} a resolution
was proposed for the case of QED. It was also described there
how the analysis of Kugo-Ojima may hold locally.
One may expect more technical 
problems in the construction of the BRST charge in the case non-abelian 
gauge theories where the free non-interacting Noether current includes 
two quantum fields. 
However, at least for the implementation of the symmetry
transformations in correlation functions, such an explicit construction of the 
BRST charge is not necessary,
as we have shown. Symmetries are implemented 
with the help of Noether currents only.

\section{Discussion}

We have presented a general method for
constructing perturbative quantum field theories with 
global and/or local symmetries. The analysis was performed
in the Bogoliubov-Shirkov-Epstein-Glaser approach.
In this framework the perturbative $S$-matrix 
is directly constructed in the asymptotic Fock 
space with only input causality and Poincar\'{e}
invariance. The construction directly yields
a finite perturbative expansion without the 
need of intermediate regularization.
The invariance of the theory under a given 
symmetry is imposed by requiring that the 
asymptotic Noether current is conserved 
at the quantum level. 

The novel feature of the present discussion 
with respect to the usual approach 
is that our results are manifestly scheme independent.
In addition, in the conventional 
approach one implicitly assumes the naive 
adiabatic limit. Our construction is 
done before the adiabatic limit is taken.
The difference between the two approaches 
is mostly seen when the symmetry condition 
is expressed in terms of the interacting 
Noether current. If the interacting current 
generates non-linear symmetries, it is 
not conserved before the adiabatic limit 
is taken. An important example is 
pure gauge theory. In this case, 
the global symmetry is BRST symmetry.
The interacting BRST current is not 
conserved before the adiabatic limit.
Nevertheless, one may still construct 
correlation functions that satisfy the 
expected Ward identities.

In the present contribution and in \cite{paper1}
we analyzed the symmetry conditions assuming that 
there are no true tree-level or loop-level
obstructions. The algebra of the symmetry transformation 
imposes integrability conditions on the 
possible form of these obstructions \cite{WZ}.
Therefore, to analyze the question of anomalies
in the present context one would have to understand
how to implement the algebra of symmetry transformations
in this framework. 
This is expected to be encoded in multi-current correlation functions.
We will   report on this issue in a future publication 
\cite{paper2}.

The Quantum Noether Condition (\ref{cons}) or (\ref{cond3}) 
leads to specific constraints (equations (\ref{n<k}), (\ref{n>k}))
that the local normalization terms should satisfy. 
We have seen that these conditions are equivalent to the 
condition that one has an invariant action.
So, one may infer the most general solution of equations
(\ref{n<k}), (\ref{n>k}) from the most general solution
of the problem of finding an action invariant 
under certain symmetry transformation rules.

For the particular case of gauge theories the global symmetry 
used in the construction is BRST symmetry. In EG one always works 
with a gauged fixed theory since one needs to have 
propagators for all fields. Therefore, the symmetry 
transformation rules are the gauged fixed ones. Physics, however,
should not depend on the particular gauge fixing chosen.
The precise connection between the results of the gauge invariant 
cohomology (which may be derived with the help of the
antifield formalism\cite{BV,HT}) and 
the present gauged-fixed formulation will 
be presented elsewhere \cite{HHS}.

The symmetry condition we proposed involves the (Lorentz invariant)
condition of conservation of the Noether current.
There are cases, however, where one has a charge that generates 
the symmetry but not a Noether current (for this to happen
the theory should not possess a Lagrangian).
A more fundamental formulation that will also cover these 
cases may be to demand that the charge that generates the symmetry 
is conserved at the quantum level (i.e. inside correlation 
functions). A precise formulation of this condition
may require a Hamiltonian reformulation 
of the EG approach. Such a reformulation may be interesting on its
own right. 

\section*{Acknowledgements}
We thank Klaus Fredenhagen and Raymond Stora for discussions. 
KS is supported by the Netherlands Organization for Scientific 
Research (NWO).

\appendix
\section{Appendix}

\renewcommand{\theequation}{A.\arabic{equation}}
\setcounter{equation}{0}

In this appendix we give the basic conventions and formulae
of the causal framework, in particular the definition 
of the interacting field.  A self-contained introduction 
to the EG construction may be found in section 3 of \cite{paper1}.
For further technical details we refer  
the reader to the literature \cite{EG0,stora,Scharf95,H95}.

We describe the construction for the case of a massive scalar 
field. The very starting point is the Fock
space ${\cal F}$ of the massive scalar field (based on a
representation space $H_s^{m}$ of the Poincar\'{e} group)
with the defining equations
\begin{equation}
(\Box + m^2) \varphi  = 0  \quad {\bf (a) } ,\quad 
[\varphi (x),\varphi (y)] = i \hbar D_{m} (x-y) \quad  {\bf (b)},
\end{equation}
where $D_{m} (x-y)= \frac{-i}{(2\pi)^3} \int dk^4 \delta(k^2-m^2) 
\sgn(k^0) \exp(-ikx)$ is the Pauli-Jordan distribution.
In contrast to the Lagrangian approach, the $S$-matrix is directly
constructed in this Fock space in the form of a formal power series
\begin{equation}
\label{GC}
 S(g) = 1 + \sum_{n=1}^\infty \frac{1}{n !} 
\int dx_1^4 \cdots dx_n^4 \quad 
T_n(x_1, \cdots, x_n; \hbar) \quad  g(x_1) \cdots g(x_n).
\end{equation}
The coupling constant $g$ is replaced by a tempered test 
function $g(x) \in {\cal S}$ 
(i.e. a smooth function rapidly decreasing at infinity) 
which switches on the interaction. 

The central objects are the $n$-point operator-valued 
distributions $T_n \in {\cal S'}$, where ${\cal S'}$
 denotes the space of functionals on ${\cal S}$. 
They should be viewed as mathematically
well-defined (renormalized) time-ordered products,
\begin{equation}
 T_n(x_1, \cdots, x_n; \hbar) = T \left[T_1(x_1) \cdots T_1(x_n)\right],
\end{equation}
of a given specific coupling, say 
$T_1={i \over \hbar} :\Phi^4: \quad {\bf (c)},$ 
which is the third defining equation in order to
specify the theory in this formalism. 
Notice that the expansion in (\ref{GC}) is {\it not} a loop expansion.
Each $T_n$ in (\ref{GC}) can receive tree-graph and loop-contributions.
One can distinguish the various contributions from the power of 
$\hbar$ that multiplies them.

Epstein and Glaser present an
explicit inductive construction of the most general perturbation series
in the sense of (\ref{GC}) which is compatible with the fundamental
axioms of relativistic quantum field theory, causality and 
Poincar\'{e} invariance.

The main guiding principle is the property  of causal 
factorization which can be stated as follows:\\
$\bullet$ Let $g_1$ and $g_2$ be two tempered test functions. Then
causal factorization means that
\begin{equation}
\label{GD}
S(g_1 + g_2) = S(g_2)  S(g_1)    \quad if \quad 
\mbox{supp} g_1 \preceq \mbox{supp} g_2
\end{equation}
the latter notion means that the support of $g_1$ and the support of
$g_2$, two closed subsets of ${\bf R}^4$, can be separated by a space
like surface.

It is  well-known that the heuristic solution for (\ref{GD}), namely
\begin{equation}
\label{heuristic}
T_n(x_1, \ldots, x_n; \hbar) = \sum_{\pi} T_1(x_{\pi (1)} ) \ldots 
T_1(x_{\pi (n)} )  \Theta (x_{\pi (1)}^0 - x_{\pi (2)}^0 ) \ldots 
\Theta (x_{\pi (n-1)}^0 - x_{\pi (n)}^0),
\end{equation}
is, in general, affected by 
ultra-violet divergences ($\pi$  runs over all permutations of 
$1, \ldots, n$).
The reason for this is that the
product of the discontinuous $\Theta$-step function with Wick monomials
like $T_1$ which are operator-valued distributions is ill-defined. 
One can handle this problem by using the usual regularization and
renormalization procedures and finally end up with the 
renormalized time-ordered products of the couplings $T_1$.

Epstein and Glaser suggest another path which leads directly to 
well-defined $T$-products without any intermediate modification of
the theory using the fundamental property of causality (\ref{GD}) as a
guide. They translate the condition (\ref{GD}) into an induction
hypothesis, $H_m, m<n$, for the $T_m$-distribution which reads 
\be \label{caus1}
H_m:\left\{
\ba{c}
\hspace{0.1cm} T_m(X \cup Y) = T_{m_1} (X) \ T_{m-m_1} (Y) \quad {\rm if}  
\quad X \succeq  Y,\quad X,Y \neq \emptyset, \ 0<m_1<m \\
\hspace{0.1cm} [T_{m_1} (X), T_{m_2} (Y)] = 0 \ {\rm if} \ X \sim Y \
(\Leftrightarrow X \succeq Y \wedge X \preceq  Y) \ \forall m_1,m_2 \le m
\ea
\right.
\ee
Here we use the short-hand notation $ T_m(x_1, \ldots, x_m; \hbar) = T(X)$;
$ \mid X \mid = m $.\\
Besides other properties they also include the Wick formula for the
$T_m$ distributions into the induction hypothesis. 
This is most easily done by including the so-called Wick submonomials  
of the specific coupling $T_1 =(i/\hbar) :\Phi^4:$ as
additional couplings in the construction 
$T_1^j := (i/\hbar) (4!/(4-j)!) :\Phi^{4-j}:, 0<j<4$.
Then the Wick formula for the $T_n$ products can be written as
\be
T_m [ T^{j_1}_1(x_1) \cdots T^{j_m}_1(x_m)]
=\!\!\!\!\sum_{s_1,..,s_m}\<0\mid T[ T^{j_1+s_1}_1(x_1)\cdots 
T^{j_m+s_m}_1(x_m)]   \mid 0\>
:\prod_{i=1}^m[\frac{\Phi^{s_i}(x_1)} {s_i !}]: 
\ee 
That such a quantity is a well-defined operator-valued distribution in
Fock space is assured by distribution theory (see Theorem O in
\cite{EG0}, 2. p. 229).  Note also that the coefficients in the Wick expansion are
now represented as
vacuum expectation values of operators. 

Now let us assume that $T_m$ distributions with all required
properties are successfully constructed for all $m<n$.
Epstein and Glaser introduce then the retarded and the
advanced $n$-point distributions (from now on 
we suppress the $\hbar$ factor in our notation):
\begin{equation}
\label{ret}
R_n(x_1,\ldots ,x_n)=T_n(x_1,\ldots ,x_n)+R'_n, 
\quad R'_n=\sum_{P_2} T_{n-n_1}(Y,x_n)\tilde{T}_{n_1}(X) 
\end{equation}
\begin{equation}
\label{adva}
A_n(x_1,\ldots ,x_n)=T_n(x_1,\ldots ,x_n)+A'_n, 
\quad A'_n=\sum_{P_2} \tilde{T}_{n_1}(X)T_{n-n_1}(Y,x_n).
\end{equation}
The sum runs over all partitions $P_2:\{x_1,\ldots x_{n-1} \}=X \cup
Y, \quad X \not= \emptyset$ into disjoint subsets with $\mid X \mid
=n_1 \ge 1, \mid Y \mid \le n-2.$
The $\tilde{T}$ are the operator-valued distributions of the inverse $S$-matrix:
\begin{equation}
S(g)^{-1}=1+\sum_{n=1}^\infty \frac {1}{n!} \int d^4x_{1}\ldots 
d^4 x_n \tilde{T}_n(x_1,\ldots x_n)g(x_1)\ldots g(x_n)
\end{equation}
The distributions $\tilde{T}$\
can be computed by formal inversion of $S(g)$:
\begin{equation}
S(g)^{-1}=(\bf 1 \rm + T)^{-1}=\bf 1\rm + \sum_{n=1}^\infty (- T)^r
\end{equation}

\begin{equation}
\tilde{T}_n(X)=\sum_{r=1}^n (-)^{r} \sum_{P_r}T_{n_1}(X_1)
\ldots T_{n_r}(X_r),
\end{equation}
where the second sum runs over all partitions $P_r$ of X into r disjoint 
subsets $X=X_1\cup\ldots\cup X_r,\quad X_j\not=\emptyset,\quad 
\mid X_j \mid =n_j.$

We stress the fact that all products of distributions are well-defined 
because the arguments are disjoint sets of points so that 
the products are tensor products of distributions.
We also remark that both sums, $R'_n$ and $A'_n$,  in contrast to $T_n$, 
contain $T_j$'s with $j \le n-1$ only
and are therefore known quantities in the inductive step from $n-1$ to
$n$. Note that the last argument $x_n$ is marked as the reference point for
the support of $R_n$ and $A_n$.
The following crucial support property is a consequence of the
causality conditions (\ref{caus1}):
\begin{equation} \label{support1}
\mbox{supp} R_m(x_1,\ldots ,x_m) \subseteq \Gamma_{m-1}^+(x_m), \quad m < n 
\end{equation}
where $\Gamma_{m-1}^+$ is the ($m-1$)-dimensional closed forward cone,
\begin{equation}
\Gamma_{m-1}^+(x_m)=\{(x_1,\ldots ,x_{m-1}) 
\mid (x_j - x_m)^2 \ge 0, x_j^0 \ge x_m^0, \forall j \}.
\end{equation}
In the difference
\begin{equation}
\label{D-dist}
D_n(x_1, \ldots ,x_n) \=d  R'_n-A'_n
\end{equation}
the unknown $n$-point distribution $T_n$ cancels. Hence this quantity
is also known in the inductive step.  With the help of the causality
conditions (\ref{caus1}) again, one shows that $D_n$
has causal support
\begin{equation}
\mbox{supp} D_n \subseteq \Gamma_{n-1}^+(x_n) \cup \Gamma_{n-1}^-(x_n) 
\end{equation}
Thus, this crucial support property is preserved in the 
inductive step from $n-1$ to $n$. 

Given this fact, the following inductive construction of the $n$-point
distribution $T_n$ becomes possible: Starting off with the known
$T_m(x_1, \ldots , x_n)$, $m \le n-1,$ one computes $A'_n, R'_n$
and $D_n = R'_n - A'_n.$ With regard to the supports, one can
decompose $D_n$ in the following way:
\begin{equation}
D_n(x_1, \ldots , x_n) = R_n (x_1, \ldots , x_n) - 
A_n (x_1, \ldots , x_n)
\end{equation}
\begin{equation}
\mbox{supp} R_n \subseteq \Gamma_{n-1}^+(x_n), \quad \mbox{supp} A_n 
\subseteq \Gamma_{n-1}^-(x_n)
\end{equation}
Having obtained these quantities we define $T'_n$ as
\begin{equation}
\label{TTT}
T'_n = R_n - R'_n = A_n - A'_n
\end{equation}
Symmetrizing over the marked variable $x_n$, we finally obtain 
the desired $T_n$,
\begin{equation}
T_n(x_1,\ldots x_n)=\sum_{\pi} \frac{1}{n!} 
T'_n(x_{\pi (1)}, \ldots x_{\pi (n)}) 
\end{equation}
One can verify that the $T_n$ satisfy the conditions (\ref{caus1}) 
and all other further properties of the induction
hypothesis \cite{EG0}.

Summing up, with the help of the corresponding causal factorization  
property of 
the $T_m$-distribution one is able to reduce the problem of 
constructing well-defined time-ordered products to the following 
splitting problem of distributions:

Given an operator-valued tempered 
distribution  $D_n \in  \cal S' ({\bf R^{4n}})$ 
with causal
support,
\begin{equation}
\mbox{supp} D_n \subseteq \Gamma_{n-1}^+(x_n) \cup \Gamma_{n-1}^-(x_n). 
\end{equation}
one has to find a pair (R, A) of
tempered distributions on ${\bf R^{4n}}$ with the following
characteristics:
\begin{equation}
\bullet\quad R, A \in  \cal S \it' (\bf R^{4n})\qquad \mbox{\bf (A)}
\end{equation}
\begin{equation}
\bullet\quad \mbox{supp} R \subset \Gamma^+(x_n), \quad \mbox{supp} A 
\subset \Gamma^-(x_n)\qquad \mbox{\bf (B)}
\end{equation}
\begin{equation}
\bullet\quad R - A = D\qquad \mbox{\bf (C)}
\end{equation}

A general solution of this problem was given by the mathematician
Malgrange some time ago \cite{Malgrange}.
As mentioned already, every
renormalization scheme solves this problem implicitly. 
The advantage of the Epstein-Glaser formulation is that it separates 
the purely technical details (which are 
essential for explicit calculations) from the 
simple physical structure of the theory.

The singular behavior of the distributions $d_n$
for $x \rightarrow 0$ is crucial for the splitting problem because 
$\Gamma^+_{n-1} (0) \cap \Gamma^-_{n-1} (0) = \{ 0 \}$.  One therefore has
to classify the singularities of distributions in this region. This
can be characterized in terms 
of the singular order $\omega$ of the 
distribution under consideration    which turns out to be
identical with the usual
power-counting degree. For details on the theory of
distribution splitting we refer to the literature
\cite{EG0,Scharf95}.

One has to ask whether the splitting solution of a
given numerical distribution $d$ with singular order $\omega(d)$ is
unique. Let $r_1 \in \cal S'$ and $r_2 \in \cal S'$ be two splitting
solutions of the given distribution $d \in \cal S'$. By
construction $r_1$ and $r_2$ have their support in $\Gamma^+$ and
agree with $d$ on $\Gamma^+\setminus \{0\}$, from which follows that $
(r_1 - r_2)$ is a tempered distribution with point support and with
singular order $\omega \le \omega(d):$ \be \mbox{supp}(r_1 - r_2)
\subset \{0\},\quad \omega(r_1-r_2) = \omega(d), \quad (r_1-r_2) \in
\cal S' \ee According to a well-known theorem in the theory of
distributions, we have 
\be \label{freeconstants}
r_1 - r_2 = \sum_{\mid a \mid=0}^{\omega_0} C_a \pa^a\delta(x).  
\ee 
In the case $\omega(d)<0$ which means that $d_n$ is regular at the
zero point, the splitting solution is thus unique.  In the case
$\omega(d)\ge 0$ the splitting solution is only determined up to a
local distribution with a fixed maximal singular degree $\omega_0 = 
\omega(d)$. The demands of causality (\ref{GD}) and translational
invariance leave the constants $ C_a$ in
(\ref{freeconstants}) undetermined. They have to be fixed by
additional normalization conditions.

One shows that, besides this 
normalization ambiguity, the $T_n$ distributions are already
fixed at all orders  by the 
fundamental axioms of QFT and the defining equations 
of the specific theory under consideration which includes the 
definition of the specific coupling $T_1$.

Having constructed the most general $S$-matrix one can
construct interacting field operators (compatible with causality and
Poincar\'{e} invariance) (second reference in \cite{EG0}, section 8) 
as follows:

One starts with an extended first order $S$-matrix
\begin{equation}
S(g,g_1,g_2, \ldots) = \int d^4 x \{T_1(x)g(x) + {i \over \hbar}( 
\Phi_1(x)g_1(x) 
+ \Phi_2(x)g_2(x)+ \ldots) \}
\end{equation}
where $\Phi_i$ represent certain Wick monomials like
$\varphi$ or $:\varphi^3:$.
Following Bogoliubov and Shirkov \cite{BS}, Epstein and Glaser defined 
the corresponding interacting fields $\Phi_i^{int}$ as functional
derivatives of the extended $S$-matrix:
\begin{equation}
\Phi_{i}^{int}(g,x) = {\hbar \over i}
S^{-1}(g,g_1, \ldots) \frac{\delta S(g,g_1,\ldots)}{ \delta g_i(x)} 
\Big|_{g_i=0} 
\end{equation}
One shows that the perturbation series for the interacting
fields is given by the advanced distributions of the corresponding
expansion of the $S$-matrix, namely
\begin{equation} \label{defint}
\Phi_i^{int}(g,x) = \Phi_i(x) + \sum_{n=1}^\infty \frac{1}{n!} 
\int d^4 x_1 \ldots d^4 x_n  
A_{n+1/n+1} (x_1, \ldots, x_n; x),
\end{equation}
where $A_{n+1/n+1}$ denotes the advanced distributions with $n$
original vertices $T_1$ and one vertex $\Phi_i$ at the $(n+1)$th
position; symbolically we may write:
\begin{equation}
A_{n+1/n+1} (x_1, \ldots, x_n; x) =  
Ad_{n+1} \left[ T_1 (x_1) \ldots T_1(x_n); \Phi_i (x) \right]
\end{equation}
One shows that the perturbative defined object $\Phi_i^{int}$
fulfills the properties like locality and  field equations
in the sense of formal power series. The definition can be regarded as
a direct construction of renormalized composite operators. Epstein
and Glaser showed that the adiabatic limit $g \rightarrow 1$ exists only
in the weak sense of expectation values in massive theories.
The limit possesses all the expected properties of a Green's function
such as causality, Lorentz covariance and the spectral condition. 
\\

\end{document}